# Teacher Noticing and Shifting of Student Epistemic Framing

Christopher A. F. Hass, Kansas State University, Manhattan, KS 66506, chris.hass@shaw.ca
Qing X. Ryan, California State Polytechnic University, Pomona, CA 91768, qxryan@cpp.edu
Eleanor C. Sayre, Kansas State University, Manhattan, KS 66506, esayre@gmail.com

**Abstract:** Through a case study, we demonstrate that teachers can tacitly notice student epistemic framing, and that this noticing can prompt instructor action which tips student framing. Our study is based on video data taken during tutorial sections and oral exams in an upper-division, undergraduate physics course. We analyze one interaction between a student and the course instructor through the lenses of teacher noticing and epistemic framing.

## Introduction

We use the theories of epistemic framing and teacher noticing to ground a case study of a student-instructor interaction from advanced undergraduate physics. A person's epistemic framing is their tacit understanding of what knowledge is useful in their current context. Following Sherin et al. (2011) we define teacher noticing as a three step process in which the teacher perceives something, interprets it, and formulates an intended response.

In our case study, we examine an instance of student-instructor interaction during an oral exam in an upper-division electromagnetism course (a core course for physics majors). During this episode the instructor notices aspects of the student's work and infers the student's framing from these. Examination of the instructor's action and the student's framing before and after the interaction allows us to connect the instructor's action to a shift in the student's framing. In doing this we demonstrate that instructors can tacitly notice and respond to student framing.

## Theory

The CAMP framework presented by Chari et al (2017) allows us to characterize student's thinking in upper division physics problem solving. This framework includes four frames: Conceptual Physics (CP), Algorithmic Physics (AP), Conceptual Mathematics (CM), and Algorithmic Mathematics (AM). The CAMP frames reflect whether students are using concepts or algorithms and whether they are using math knowledge or physics knowledge. To effectively solve a problem, students must move through a series of frames.

Russ & Luna (2013) connect teachers' epistemic framing to local patterns in teacher noticing. Given that instructors interact with students during problem solving activities, and that they notice many features of student problem solving, we suspect that they may also tacitly notice student framing. Also, previous work indicates that student framing before and after interaction with instructors is often different (Chari et al. 2017). By employing the CAMP framework, and stimulated recall techniques similar to Russ and Luna we demonstrate that instructors can tacitly notice and act to change (tip) student framing.

## Methods and Context

We examined video data from a post-secondary, upper-division physics course on electromagnetism. The full data set includes video of in-class problem solving activities, and video of student oral exams. Our poster focuses on an oral exam since it has frequent and sustained instructor interaction. We screened videos for interactions, and then coded our selected interaction using an iteratively refined codebook developed by two of the authors. Our case study consists of a 4 minute period out of a 37 minute exam. We chose this interaction because it is simple, and clearly shows the relation between instructor noticing and student frame tipping. We recognize that oral exams present the simplest instructor-student system (one instructor, one student, concentrated interaction) that isn't present in other classroom interactions; however, in this study we seek only to show that tacit noticing of frames is possible, so we chose data where it was most likely to occur.

Our focal student, "Abbey" is a 3rd year physics undergraduate student, taking an upper division electromagnetism (E&M) course. Abbey is chosen because she vocalizes her thoughts well, writes neatly on the whiteboard, and reasons clearly. In her oral exam she is given a physics problem to solve on the whiteboard with the instructor present. Abbey is allowed to bring a ``cheat sheet" into the exam. Abbey created her own cheat sheet, and neither we nor the instructor have access to it. During the exam, the instructor occasionally asks questions about Abbey's work, and provides small prompts or guidance when Abbey gets stuck.

In this particular example, Abbey is asked to find the electric field (E) of an infinite, uniformly charged sheet. This is a typical problem in introductory E&M courses. It has a standard solution using Gauss's flux law, which says the integral of electric field flux through a closed surface is proportional to the enclosed charge.

During her solution, Abbey sets up the geometry of her solution in a way which does not take advantage of the symmetry of the charge distribution, eliminating the calculational advantage of the Gauss's flux law treatment and rendering the solution much more mathematically cumbersome. It's worth noting here that Abbey's geometric set up is appropriate when solving a different but related problem, the electric field from an infinite line of charge. This case study analyzes the instructor noticing and responding to this issue.

## Analysis

The instructor begins the oral exam by stating the problem and drawing a diagram. After taking a moment to consider the problem, Abbey says "It's kind of like, it's, well, Gauss's Law", identifying the physics algorithm she wishes to use. Following the process, she recalls and records a physical law, and drawing on the diagram she says "I'm just going to draw a cylinder, because I like cylinders". While recognizing the algorithm requires her to enclose the electric charge with a shape, Abbey does not pause to consider how geometry affects the choice of shape. She frequently copies from her "cheat sheet" during this time; appearing to follow a process written there. Abbey spends her time here following a set of step by step instructions for calculating the electric field from a physical law; thus we interpret that Abbey is in the Algorithmic Physics (AP) frame.

As Abbey draws her cylinder (Figure 2, right side), the instructor notices that the cylinder she has chosen is going to complicate the solution. The instructor asks, "What is the direction of [the electric field]?". In the stimulated recall interview, the instructor told us that Abbey was just following a set of steps. The instructor told us they wanted Abbey to think about how the electric field direction should influence the orientation of the cylinder. This implies for us that the instructor tacitly noticed that Abbey was in the algorithmic physics frame.

In response to the instructor's question, Abbey's statements change. She makes statements such as "The direction of E... I'm going to say, perpendicular to the sheet", where she is thinking about the physical behavior of the electric field. When asked to elaborate she says "I remember, if you have a line of charge [...] then all of the horizontal components cancel [...]", and clarifies when asked by the instructor how this is similar to the plane, "If you think about [the components in the plane] they all cancel, and you're just left with the radially outward components", she argues you are left with only components out of the plane. Rather than following steps, these are statements about Abbey's conceptual understanding of how the geometry of a system affects its physical behavior. We conclude that Abbey's framing is now Conceptual Physics.

We recognize that this study is limited. Teacher noticing is situated, and we cannot be certain what tacit noticing of framing may look like a classroom environment based on this study. However, our purpose was merely to show that there are instances where teachers tacitly notice and tip student frames. Future work may generalize this idea to classroom environments.

## Conclusion

Existing research shows that student-instructor interactions have the potential to change the way students think while solving problems. In this paper we have shown that it is possible for teachers to tip student framing in response to their tacit noticing of student framing. Extending this work in the future, we will examine the types of actions that instructor noticing prompts, and the effects those actions have on student framing, especially in more complicated classroom scenarios. Using this analysis we hope to develop supporting materials to help instructors be more sensitive to student framing and more effective in interacting with students.

## Acknowledgements

Thanks to Amogh Sirnoorkar, Mary Bridget Kustusch, Scott Franklin, Grace Heath, Alana Uriarte, Darwin Del Agunos, Manny Gomez-Bera, and Kyle Benjamin for their contributions. This work was supported by the National Science Foundation under Grant No(s). DUE1726479/1726113, REU-17547477, and DUE-1430967. Paid in part by: Kansas State University, Dep. of Physics; California State Polytechnic University, Pomona, Dep. of Physics; Rochester Institute of Technology Dep. of Physics; DePaul University, Dep. of Physics